\documentclass[epj,nopacs,final]{svjour}
\usepackage{graphicx}
\usepackage{amssymb}
\usepackage{amsmath}

\begin{document}
\title{Thermodynamic anomalies in the presence of dissipation:
       from the free particle to the harmonic oscillator}
\author{Robert Adamietz\inst{1} \and Gert-Ludwig Ingold\inst{1}\fnmsep%
        \thanks{\email{gert.ingold@physik.uni-augsburg.de}}
        \and Ulrich Weiss\inst{2}}
\institute{Institut f\"ur Physik, Universit\"at Augsburg, D-86135 Augsburg \and
           II. Institut f\"ur Theoretische Physik, Universit\"at Stuttgart,
             D-70550 Stuttgart}
\authorrunning{R. Adamietz et al.}
\titlerunning{Thermodynamic anomalies in the presence of dissipation}

\abstract{%
A free particle coupled to a heat bath can exhibit a number of thermodynamic
anomalies like a negative specific heat or reentrant classicality. These
low-temperature phenomena are expected to be modified at very low temperatures
where finite-size effects associated with the discreteness of the energy
spectrum become relevant. In this paper, we explore in which form the
thermodynamic anomalies of the free damped particle appear for a damped
harmonic oscillator. Since the discreteness of the oscillator's energy spectrum
is fully accounted for, the results are valid for arbitrary temperatures.  As
expected, they are in agreement with the third law of thermodynamics and
indicate how the thermodynamic anomalies of the free damped particle can be
reconciled with the third law. Particular attention is paid to the transition
from the harmonic oscillator to the free particle when the limit of the
oscillator frequency to zero is taken.}

\maketitle

\section{Introduction}

The thermodynamic properties of a free particle coupled to a heat bath can
exhibit several interesting effects. While the thermodynamic properties of an
isolated free particle remain classical for arbitrary temperatures, the
coupling to an Ohmic environment can provide a mechanism to bring the specific
heat of the free damped particle down to zero in the zero-temperature limit.
Thus, the validity of the third law of thermodynamics in this specific case is
ensured by the particle's environment \cite{hangg06}. For sufficiently strong
coupling, the specific heat obtained from the reduced partition function can
even become negative \cite{hangg08}. This phenomenon may be understood in terms
of a modification of the density of states of the heat bath caused by level
repulsion due to coupling to the free particle \cite{ingol12}. Negative
specific heats have also been discussed in the context of Kondo superconductors
\cite{zitko09}, quantum impurity systems \cite{merke12}, XY spin chains
\cite{campi10}, two-level fluctuators \cite{campi09}, and energy transport in
proteins \cite{sulai10}.

Recently, it was found that the vanishing specific heat in the
zero-temperature limit is specific to Ohmic heat baths \cite{spren13}. Subohmic
baths, i.e.\ baths with an increased density of low-frequency degrees of
freedom with respect to an Ohmic bath, can even give rise to negative specific
heats in the zero-temperature limit. In such situations, finite-size effects
must be taken into account in order to save the third law of thermodynamics.

Another interesting phenomenon arises for sufficiently superohmic environments
where the bath spectral density is significantly suppressed at low frequencies. 
With decreasing temperature, the specific heat decreases under the influence
of the environmental coupling. However, at even lower temperatures, the specific
heat rises again up to its classical value \cite{spren13}. The low density of
low-frequency bath modes renders the bath inefficient in decreasing the specific
heat. Here, again finite-size effects need to be incorporated in order to obtain
a correct description of the thermodynamic properties at extremely low 
temperatures.

With the notable exception of an Ohmic environment, dissipation is not
sufficient to guarantee the validity of the third law of thermodynamics for the
free particle. In general, finite-size effects due to placing the particle into
a box need to be considered. They are expected to become relevant when the
temperature drops below a value related to the ground state energy of the
particle confined to the box. However, this temperature scale can be made
arbitrarily small by making the box sufficiently large. While the finite-size
effects then play their role in ensuring the validity of the third law, they do
not necessarily impede the observation of the thermodynamic anomalies discussed
above.

Evaluating the thermodynamic properties of a damped particle in a box is a
complicated task which, in general, needs to be done numerically. Perturbative
approaches are not sufficient for our purpose because the anomalies of interest
here occur at relatively strong damping. Numerically, a damped particle on a
finite chain has been treated, but there the focus was on the dissipative phase
transition \cite{sabio08}.

An analytically more tractable system is the damped harmonic oscillator for which
the discreteness of the system's energy spectrum is inherent. Some results for
thermodynamic low-temperature properties in the presence of non-Ohmic damping
have been discussed in \cite{wang12}. The related problem of a charged particle
in a magnetic field and a harmonic potential has been widely studied
\cite{bandy09,kumar09,datta10,bandy10,bandy10a,kumar13,kumar14}. Also, systems
containing more than one harmonic degree of freedom have been discussed recently
\cite{haseg11}.

In the following, we will focus on the properties of the damped harmonic
oscillator for small oscillator frequency. This will allow us to make connection
to the free particle and in particular to decide, whether the thermodynamic
anomalies found for the free particle are accessible to observation in a harmonic
potential. We will start in Sects.~\ref{sec:reduced_partition_function} and
\ref{sec:gammahat} by reviewing some basic relations for damped quantum systems.
Specifically, in Sect.~\ref{sec:reduced_partition_function} we introduce the
reduced partition function on which our evaluation of the thermodynamic
quantities will be based. The explicit expressions for the reduced partition
functions will give us a first idea of the transition from the damped
harmonic oscillator to the damped free particle. In Sect.~\ref{sec:gammahat}
we will introduce the Laplace transform of the damping kernel which within the
scope of this paper is the central quantity describing the heat bath.

In Sect.~\ref{sec:change_of_bath_dos} we will take the point of view of
the bath \cite{ingol12} and study how the bath density of states is modified
when a harmonic oscillator is coupled to it. The results will provide us with
important information about the relation between the damped harmonic oscillator
and the damped free particle. In Sects.~\ref{sec:specific_heat} and \ref{sec:entropy},
we will discuss the specific heat and entropy, respectively, and we will make
the connection between features found in the specific heat of a free damped
particle and those found for a damped harmonic oscillator. Finally, in
Sect.~\ref{sec:conclusions}, we will present our conclusions.

\section{Reduced partition function}
\label{sec:reduced_partition_function}

As was first pointed out in Ref.~\cite{hangg06}, thermodynamic quantities
in the quantum regime beyond the weak-coupling limit between system (S) and
heat bath (B) are not uniquely defined. Here, we choose to base the derivation
of thermodynamic quantities on the reduced partition function of the system
\begin{equation}
\label{eq:reduced_partition_function}
\mathcal{Z} = \frac{\mathcal{Z}_\text{S+B}}{\mathcal{Z}_\text{B}}
\end{equation}
obtained from the partition function $\mathcal{Z}_\text{S+B}$ of system and
heat bath and the partition function $\mathcal{Z}_\text{B}$ of the heat bath
alone. In the absence of any coupling between system and bath, we have
$\mathcal{Z}_\text{S+B}=\mathcal{Z}_\text{S}\mathcal{Z}_\text{B}$ and the
reduced partition function (\ref{eq:reduced_partition_function}) equals the
partition function of the system $\mathcal{Z}_\text{S}$. In general, the
reduced partition function will differ from the partition function of the
uncoupled system, thereby describing the influence of the heat bath on the
thermodynamic properties of the system.

Applying the usual thermodynamic relations, our choice implies that the
thermodynamic quantities of a damped quantum system like the specific heat
or the entropy are actually given by the change of this quantity when the
system is coupled to the environment.

For example, the specific heat of the damped system is expressed as the
difference between the specific heat of system and bath and the specific heat of
the bath alone
\begin{equation}
\label{eq:specific_heat}
C=C_\text{S+B}-C_\text{B}\,.
\end{equation}
The specific heat, as well as other thermodynamic quantities defined in this way,
can thus be regarded as a property of the damped system and, alternatively, as a
change in the properties of the heat bath. The latter point of view will be taken
in Sect.~\ref{sec:change_of_bath_dos}.

For the free damped particle, the reduced partition function is given by
\cite{weiss12}
\begin{equation}
\label{eq:z_fp}
\mathcal{Z} = \sqrt{\frac{\pi}{4\beta E_\text{g}}}
              \prod_{n=1}^\infty\frac{\nu_n}{\nu_n+\hat\gamma(\nu_n)}
\end{equation}
where $\beta=1/k_\text{B}T$ is proportional to the inverse temperature $T$ and
the Matsubara frequencies are given by $\nu_n=2\pi n/\hbar\beta$.  The Laplace
transform $\hat\gamma(z)$ of the damping kernel will be introduced in detail
below in Sect.~\ref{sec:gammahat}. In the prefactor,
$E_\text{g}=\hbar^2\pi^2/2ML^2$ is the ground state energy of a particle of
mass $M$ in a one-dimensional box of width $L$. The finite box is merely needed
for normalization. For the purpose of this paper, the width $L$ is assumed to
be so large that the discrete level structure becomes only relevant at
temperatures much lower than those of interest here.

For a harmonic oscillator of frequency $\omega_0$, the reduced partition
function reads \cite{weiss12}
\begin{equation}
\label{eq:z_ho}
\mathcal{Z} = \frac{1}{\hbar\beta\omega_0}\prod_{n=1}^\infty\frac{\nu_n^2}
              {\nu_n^2+\nu_n\hat\gamma(\nu_n)+\omega_0^2}\,.
\end{equation}
The factors in front of the infinite product in (\ref{eq:z_fp}) and
(\ref{eq:z_ho}) determine the behavior in the classical regime. The different
powers of temperature, $1/2$ for the free particle and $1$ for the harmonic
oscillator, are a consequence of the different number of degrees of freedom.
For example, they lead to a high-temperature specific heat of $k_\text{B}/2$ for
the free particle and of $k_\text{B}$ for the harmonic oscillator.

Thus, the limit $\omega_0\to 0$ of the harmonic oscillator cannot lead to the
free particle in a smooth way. If we disregard the difference arising from the
classical factor for the moment, there remains a difference due to the infinite
products in (\ref{eq:z_fp}) and (\ref{eq:z_ho}). However, the infinite products
can only differ significantly at low temperatures $k_\text{B}T\ll
\hbar\omega_0/2\pi$ which become arbitrarily small in the limit $\omega_0\to
0$.

Already on the basis of the reduced partition functions (\ref{eq:z_fp}) and
(\ref{eq:z_ho}) we can expect differences to occur between the thermodynamic
quantities of damped free particle and damped harmonic oscillator which are
solely determined by their classical expressions. Still, these differences can
be relevant even deep into the quantum regime. In addition, below a certain
temperature, the discrete energy spectrum of the harmonic oscillator will start
to play a role and, in particular, will ensure the validity of the third law of
thermodynamics, as we shall see.

\section{Laplace transform of the damping kernel}
\label{sec:gammahat}

The heat bath is described within a Caldeira-Leggett model where the system is
coupled bilinearly through its position to the positions of a set of harmonic
bath oscillators \cite{weiss12}. However, we need not be concerned with the
details of this model because the only quantity of relevance for the system
properties is the spectral density of the system-bath coupling $J(\omega)$. It
is related to the Laplace transform of the damping kernel appearing in
(\ref{eq:z_fp}) and (\ref{eq:z_ho}) by means of
\begin{equation}
\label{eq:gammahat_j}
\hat\gamma(z) = \frac{2}{\pi M}\int_0^\infty\text{d}\omega\frac{J(\omega)}{\omega}
                \frac{z}{\omega^2+z^2}\,.
\end{equation}

To be specific, we choose a spectral density of the form
\begin{equation}
\label{eq:JofOmega}
J(\omega) =
M\gamma\omega^s\frac{\omega_\text{c}^{2p-s+1}}{(\omega_\text{c}^2+\omega^2)^p}\,.
\end{equation}
with $0<s<2p+2$. Here, $M$ is the mass associated with the system degree of
freedom and $\gamma$ determines the damping strength. At low frequencies, the
spectral density (\ref{eq:JofOmega}) increases proportional to $\omega^s$ where
the exponent $s$ is decisive for the low-temperature thermodynamics. In order
to avoid ultraviolet divergences, we have chosen a generalized Drude-type
cutoff represented by the last factor on the right-hand side of
(\ref{eq:JofOmega}). It suppresses the spectral density of bath oscillators
above a frequency scale determined by the cutoff frequency $\omega_\text{c}$.
Other choices for the cutoff are possible but are not expected to affect our
results in an important manner.

In Ref.~\cite{spren13} it was shown for a free damped particle subject to
a bath described by (\ref{eq:JofOmega}), that the specific heat approaches
$C/k_\text{B}=(s-1)/2$ for $s\leq 2$ as the temperature is lowered towards zero.
The particular case $s=1$ corresponds to Ohmic damping where the specific heat
of the free damped particle vanishes at zero temperature. For subohmic damping,
$s<1$, the specific heat even tends to a negative value in the low-temperature
limit as long as finite-size effects remain irrelevant. For superohmic damping
with $s\geq2$, the classical value  $C/k_\text{B}=1/2$ is approached in the
low-temperature regime.

Employing the relation (\ref{eq:gammahat_j}), one finds that for the spectral
density (\ref{eq:JofOmega}), the Laplace transform of the damping kernel can be
expressed in terms of a hypergeometric function \cite{spren13}. For our
purposes, it will be sufficient to consider integer values for the exponent
$p$ appearing in the cutoff function. Then, the Laplace transform of the
damping kernel can be expressed in terms of a finite sum as
\begin{equation}
\label{eq:gammahat_int_p}
\begin{aligned}
\hat\gamma(z) &= \frac{\gamma\omega_\text{c}^{2p}}{(\omega_\text{c}^2-z^2)^p}
   \left[\frac{(z/\omega_\text{c})^{s-1}}{\sin(\frac{\pi s}{2})}\right.\\
   &\; \left.+\frac{1}{\pi}\sum_{n=1}^{p}\frac{(-1)^{n}}{n-\frac{s}{2}}
         \frac{B(\frac{s}{2}, p+1-\frac{s}{2})}{B(n, p-n+1)}
         \left(\frac{z}{\omega_\text{c}}\right)^{2n-1}\right]\,.
\end{aligned}
\end{equation}
The beta function is defined in terms of gamma functions as $B(x,
y)=\Gamma(x)\Gamma(y)/\Gamma(x+y)$. The expression (\ref{eq:gammahat_int_p}) is
not valid for even integer values of $s$ where logarithmic terms occur. The
expression for $s=2$ can be found in eq.~(15) of Ref.~\cite{spren13}.

We will mostly be interested in the Ohmic case $s=1$ and the superohmic case
$s=3$ where analytical results can be obtained and which allow to discuss the
main thermodynamic anomalies. For $s=1$ and sufficiently strong damping, the
free damped particle has a negative specific heat at low temperatures
\cite{hangg08} while for $s=3$ reentrant classicality is observed \cite{spren13}.

For odd integer $s$, the expression (\ref{eq:gammahat_int_p}) can be written as
\begin{equation} \label{eq:gamma_poly}
\hat\gamma(z) = \frac{P(z;s,p)}{(\omega_\text{c}+z)^p}\,,   
\end{equation}
where $P(z;s,p)$ is a polynomial in $z$ of order $z^{p-1}$, in which the term
of zeroth order is absent for $s=3, 5, \ldots$  Particular cases are
\begin{align}  \nonumber
P(z;1,1) &= \gamma\omega_\text{c} \,  \\
P(z;1,2) &= \gamma\omega_\text{c}\left(\omega_\text{c} +\frac{z}{2}\right) \, \label{eq:cases_spec} \\
P(z;3,2) &=  \gamma\omega_\text{c}  \frac{z}{2}  \;  .  \nonumber
\end{align}

\section{Change of bath density of states}
\label{sec:change_of_bath_dos}

As mentioned above, the specific heat (\ref{eq:specific_heat}) can either be
viewed as specific heat of the system modified by the coupling to the heat bath
or as change in the specific heat of the bath when the system degree of freedom
is coupled to it. The latter point of view was taken in
Ref.~\cite{ingol12} where it was shown that the specific heat
(\ref{eq:specific_heat}) of the damped free particle can be expressed in terms
of the change in the bath density of states together with the well-known
expression for the specific heat of a harmonic oscillator.

The change in the bath density of states is defined as
\begin{equation}
\label{eq:xi}
\xi(\omega) = \sum_n[\delta(\omega-\omega_n)-\delta(\omega-\omega^0_n)]\,,
\end{equation}
where $\omega^0_n$ are the eigenfrequencies of the bath oscillators in the
absence of the system degree of freedom and $\omega_n$ are the frequencies of
the eigenmodes of system and bath coupled to each other. The density of states
$\xi(\omega)$ of interest here should not be confused with the spectral density
of the coupling $J(\omega)$ introduced in the context of (\ref{eq:gammahat_j})
and (\ref{eq:JofOmega}).

For the harmonic oscillator (osc) and the free particle (fp) we obtain
the change of the bath density of states as
\begin{align}
\label{eq:xi_osc}
\xi_{\rm osc}(\omega) &= \frac{1}{\pi}\text{Im}\frac{\partial
\ln[\hat{\chi}(-\text{i}\,\omega)]}{\partial\omega} \, ,    \\   \label{eq:xi_fp}
\xi_{\rm fp}(\omega) &= \frac{1}{\pi}\text{Im}\frac{\partial
\ln[\hat{\mathcal{R}}(-\text{i}\omega)]}{\partial\omega}  \,,
\end{align}
respectively.
Here, 
\begin{equation}
\label{eq:chi_osc}
\hat{\chi}(z) = \frac{1}{z^2+z\hat\gamma(z) + \omega_0^2}
\end{equation}
is the dynamical susceptibility of the damped harmonic oscillator and
\begin{equation}
\label{eq:R_fp}
\hat{\mathcal{R}}(z) = \frac{1}{z+\hat\gamma(z)}
\end{equation}
is the dynamical velocity response of the damped free particle. Further, Im
denotes the imaginary part. Observe that $\xi_{\rm osc}(\omega)$ concurs with
$\xi_{\rm fp}(\omega)$ in the limit $\omega_0\to 0$ for any non-zero frequency
$\omega$.

With the form (\ref{eq:gamma_poly}) for $\hat\gamma(z)$, the poles of
$\hat\chi(z)$ are determined by the zeros $z_i$ of the polynomial equation
\begin{equation}
\label{eq:polynomial_eq}
N(z;\omega_0) \equiv (z^2+\omega_0^2)(z+\omega_\text{c})^p+zP(z;s,p)=0 \, .
\end{equation}
Each individual pole of $\hat\chi(z)$ contributes a Lorentzian to the change of
the bath density of states. Denoting the real and imaginary part of $z$ by
$z'$ and $z''$, respectively, we then have
\begin{equation}
\label{eq:xi_lorentz}
\xi(\omega) = \frac{1}{\pi}\left(-p\frac{\omega_\text{c}}{\omega^2+\omega_\text{c}^2}
- \sum_i \frac{z_i'}{(\omega+z_i'')^2+ z_i'^{2}} \right)\, .
\end{equation}
Since  eq.~(\ref{eq:polynomial_eq}) applies to a damped system, the real part
of each individual zero is negative, $z_i'<0$.  Thus, the respective
Lorentzians give positive contributions to the density (\ref{eq:xi_lorentz}).

Consider next eq. (\ref{eq:polynomial_eq}) in the free particle limit
$\omega_0=0$. With eqs.~(\ref{eq:gammahat_int_p}) and (\ref{eq:cases_spec}) we
see that the polynomial $N(z;0)$ has a simple zero at the origin, $z=0$, for
$s=1$, and a double zero at the origin for $s=3, 5, \ldots$  These solutions
of eq.~(\ref{eq:polynomial_eq}) do not contribute to the change of the bath
density of states (\ref{eq:xi_lorentz}). As a result, for any $p$ the density
$\xi_{\rm fp}(\omega)$ has one Lorentzian less compared with $\xi_{\rm
osc}(\omega)$ for $s=1$ and two Lorentzians less for $s=3, 5, \ldots$

The absence of one or two low-frequency peaks in the density $\xi_{\rm
fp}(\omega)$ compared to $\xi_{\rm osc}(\omega)$ will lead to qualitative
differences in the thermodynamic behavior of the damped free particle and
harmonic oscillator. In this respect, the value of the difference of
the integrated densities 
\begin{equation}\label{eq:sumrule}
\begin{aligned}
\Delta\Sigma(s) &\equiv \Sigma_{\rm osc}(s)-\Sigma_{\rm fp}(s)\\
&=\int_0^\infty {\rm d}\omega\,[\, \xi_{\rm osc} (\omega;s) - \xi_{\rm fp} (\omega;s) \,]
\end{aligned}
\end{equation}
plays a crucial role, as we shall see.

The preceding discussion will now be illustrated by considering in more detail
the important cases $s=1$ and $3$. We start with Ohmic damping, $s=1$, and a
cutoff characterized by $p=1$. For the free particle, the non-vanishing zeros
of the equation $N(z;0)=0$, which contribute to the change of the spectral
density $\xi_{\rm fp}(\omega)$, are 
\begin{equation}
\label{eq:fp_1_1_z12}
z_{1, 2} = - \frac{\omega_\text{c}}{2}\left(1\pm\text{i}\sqrt{\frac{4\gamma}{\omega_\text{c}}-1}
 \right)\,.
\end{equation}
These zeros supply the density $\xi_{\rm fp}(\omega)$ with two Lorentzians.
Depending on whether the damping strength $\gamma$ is smaller or larger than
$\omega_\text{c}/4$, they are centered around zero frequency or finite
frequency with opposite signs, respectively.

Proceeding to the harmonic oscillator, we account for a small frequency
$\omega_0$. In this case, the expressions (\ref{eq:fp_1_1_z12}) receive minor
corrections of order $\omega_0^2$. The most important change, however, consists
in an additional real zero acquired by $N(z; \omega_0)$. To leading order in
$\omega_0$, this zero is independent of $p$ and located at 
\begin{equation}
\label{eq:ho_1_1_z3}
z_3 = - \frac{\omega_0^2}{\gamma}\,.
\end{equation}

The Lorentzian associated with the zero (\ref{eq:ho_1_1_z3}) is centered at
$\omega=0$ and becomes a delta function in the limit of vanishing oscillator
frequency, $\omega_0\to0$. Since this contribution is missing in the density
$\xi_{\rm fp}(\omega)$, we have the concise exact relation
\begin{equation}
\label{eq:sum_diff_1}
\Delta\Sigma(s=1) =  \frac{1}{2} \,  .
\end{equation}

\begin{figure}
 \centerline{\includegraphics[width=0.95\columnwidth]{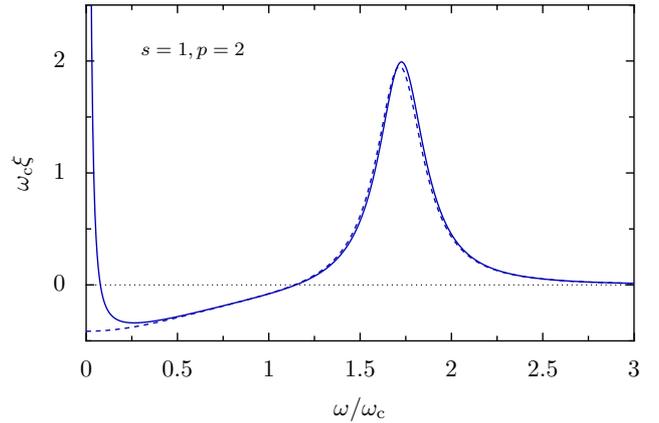}}
 \caption{Change of the density of bath oscillators $\xi_{\rm osc}(\omega)$
          for a spectral density (\ref{eq:JofOmega}) characterized by $s=1$
          and $p=2$, damping constant $\gamma=5\omega_\text{c}$, and an
          oscillator frequency $\omega_0=0.2\omega_\text{c}$ (solid line).
          For comparison, the dashed line gives the density
          $\xi_{\rm fp}(\omega)$ for the free damped particle.}
 \label{fig:changedos_1_2}
\end{figure}

Figure~\ref{fig:changedos_1_2} compares the change of the density of bath
oscillators for a damped harmonic oscillator of frequency
$\omega_0=0.2\omega_\text{c}$ (solid line) with that for a damped free particle
(dashed line) for an Ohmic spectral density, $s=1$. In contrast to the previous
analytical considerations, we now choose a cutoff characterized by $p=2$ in
order to facilitate comparison with the superohmic case to be discussed below.
Apart from the Lorentzian with negative weight given by the first term in
(\ref{eq:xi_lorentz}), the structure of $\xi_\text{fp}(\omega)$ for the free
damped particle is dominated by a peak at finite frequency determined by a
solution of $N(z; 0)=0$ analogous to (\ref{eq:fp_1_1_z12}).

The main difference between free particle and harmonic oscillator concerns the
behavior at low frequencies. Here, the real solution (\ref{eq:ho_1_1_z3}) of
(\ref{eq:polynomial_eq}) for the damped harmonic oscillator leads to a peak
centered around zero frequency and dominating the low-frequency behavior. As we
will see later on, it is this peak which is responsible for the main difference
in the low-temperature behavior of thermodynamic quantities between the free
particle and the harmonic oscillator.

Consider next the case of a superohmic bath with $s=3$ and $p=2$. With respect
to the free damped particle, eq.~(\ref{eq:polynomial_eq}) with $\omega_0=0$
again yields two nontrivial solutions,
\begin{equation}
\label{eq:fp_3_2_z12}
z_{1,2} = -\omega_\text{c}\pm\text{i}\sqrt{\frac{\gamma\omega_\text{c}}{2}}\, .
\end{equation}
These account for peaks at finite frequencies $\pm\sqrt{\gamma\omega_{\rm c}/2}$,
very much like in the Ohmic case. 

Advancing from the free particle to the harmonic oscillator, the originally
trivial double zeros $z=0$ of the equation $N(z;0)=0$ turn into  a pair of
complex solutions of the pole condition $N(z,\omega_0)=0$ with the leading real
and imaginary parts given by 
\begin{equation}
\label{eq:ho_3_2_z34}
z_{3,4} = \pm\text{i}\frac{\omega_0}{(1+\gamma/2\omega_\text{c})^{1/2}}
-\frac{\gamma}{2\omega_\text{c}^2}\frac{\omega_0^2}{(1+\gamma/2\omega_\text{c})^2}\,.
\end{equation}
In contrast to the single peak centered at the origin in the Ohmic case, there
are now two low-frequency peaks positioned symmetrically about the origin with
displacement of the order of $\omega_0$. Since these two peaks are missing in the
density $\xi_{\rm fp}(\omega)$, we now have
\begin{equation}\label{eq:sum_diff_3}
\Delta\Sigma(s=3) = 1 \, .
\end{equation}

\begin{figure}
 \centerline{\includegraphics[width=0.95\columnwidth]{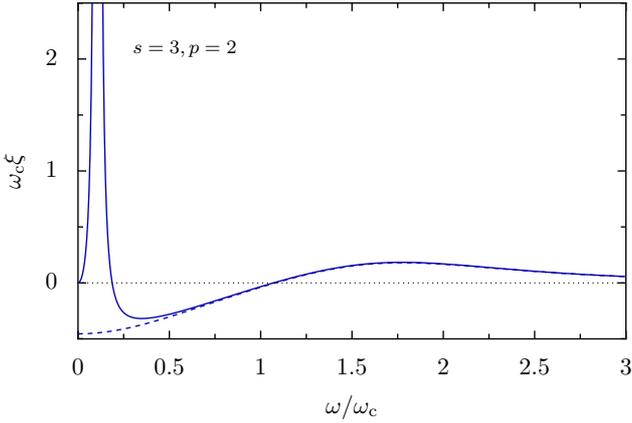}}
 \caption{Change of the density of bath oscillators $\xi_{\rm osc}(\omega)$
          for a spectral density (\ref{eq:JofOmega}) characterized by $s=3$
          and $p=2$, damping constant $\gamma=5\omega_\text{c}$, and an
          oscillator frequency $\omega_0=0.2\omega_\text{c}$ (solid line).
          For comparison, the dashed line gives the density
          $\xi_{\rm fp}(\omega)$ for the free damped particle.}
 \label{fig:changedos_3_2}
\end{figure}

The change of the density of bath oscillators $\xi(\omega)$ for the superohmic
case with $s=3$, $p=2$, and damping strength $\gamma=5\omega_\text{c}$ is
displayed in Fig.~\ref{fig:changedos_3_2}. As in Fig.~\ref{fig:changedos_1_2},
the solid line corresponds to a damped harmonic oscillator with
$\omega_0=0.2\omega_\text{c}$ while the dashed line refers to a free damped
particle. Comparison of Figs.~\ref{fig:changedos_1_2} and
\ref{fig:changedos_3_2} clearly shows the difference in the low-frequency
behavior of $\xi_\text{osc}(\omega)$ for the damped harmonic oscillator coupled
to an Ohmic and a superohmic heat bath.

A more detailed analysis of the low-frequency peak for arbitrary exponents $s$
of the spectral density of the coupling is given in appendix~\ref{sec:appa}. 
In particular, it is shown there that the results (\ref{eq:sum_diff_1}) and
(\ref{eq:sum_diff_3}) can be generalized to arbitrary $s$ as
\begin{equation}
\label{eq:sum_diff_s}
\Delta\Sigma(s) =
\begin{cases}
\dfrac{s}{2} &\text{for $0 < s < 2$}\\[0.8em]
1 &\text{for $s \geq 2$}\,.
\end{cases}
\end{equation}
We will make use of this result in the discussion of the specific heat in the
next section.

\section{Specific heat}
\label{sec:specific_heat}

Since the densities of state appearing in (\ref{eq:xi}) can be thought of being
constituted by harmonic oscillators, the specific heat defined according to
(\ref{eq:specific_heat}) can be obtained from the specific heat $C_\text{ho}$
of a single harmonic oscillator with frequency $\omega$
\begin{equation}
\label{eq:cho}
C_\text{ho} = k_\text{B}\left(\frac{\hbar\beta\omega}
                                   {2\sinh(\hbar\beta\omega/2)}\right)^2\,.
\end{equation}
This specific heat vanishes in the zero-temperature limit and approaches $k_\text{B}$
in the classical limit. Therefore, we can generally write
\begin{align}
C &= C_0+\int_0^\infty\text{d}\omega\,\xi(\omega)C_\text{ho}(\omega)\\
  \label{eq:c_xi_infty}
  &= C_\infty+\int_0^\infty\text{d}\omega \,\xi(\omega)
     [C_\text{ho}(\omega)-k_\text{B}]\,,
\end{align}
where $\xi(\omega)$ can either be the change of the density of states
(\ref{eq:xi_osc}) or (\ref{eq:xi_fp}) for the damped harmonic oscillator or the
free damped particle, respectively. $C_0$ is the specific heat of the damped
system in the zero-temperature limit while $C_\infty$ is its classical value.
The two values are related by
\begin{equation}
\label{eq:c_inf_0}
C_\infty=C_0+k_\text{B}\Sigma\,,
\end{equation}
where $\Sigma$ denotes the integrated change of the density of states in agreement
with the notation introduced in (\ref{eq:sumrule}).

For the damped harmonic oscillator, we obtain from the prefactor in the
partition function (\ref{eq:z_ho}) the classical value of the specific heat
$C_\infty=k_\text{B}$. Observing that the defining expression (\ref{eq:xi_osc})
is a partial derivative and that the real part of $\hat\gamma(-\text{i}\omega)$
is positive, we find for all $s$ the sum rule
\begin{equation}
\label{eq:sumrule_osc}
\Sigma_{\rm osc}=1\,,
\end{equation}
and thus $C_0=0$. It is therefore always guaranteed that the specific heat for
the damped harmonic oscillator vanishes at zero temperature.  The specific heat
as a function of temperature can now be expressed as
\begin{equation}
\label{eq:cdampedho}
C = \int_0^\infty\text{d}\omega\,\xi_{\rm osc}(\omega)C_\text{ho}(\omega)\,.
\end{equation}
This result can also be derived by means of $C=\partial U/\partial T$ from the
expression for the internal energy obtained by Ford \textit{et al.} for a damped
harmonic oscillator \cite{ford85,ford88}. 

For the damped free particle, the situation is a bit more complex. Again the
classical value of the specific heat can be found from the prefactor of the
corresponding partition function (\ref{eq:z_fp}) and takes the value
$C_\infty=k_\text{B}/2$ as expected. For the sum rule $\Sigma_{\rm fp}(s)$ we
have to distinguish the cases $0<s<2$ and $s\geq2$. In the first case, we find
from the expression (\ref{eq:xi_fp}) that the low-frequency behavior of the
change of the bath density of states is dominated by the first term in
(\ref{eq:gamma_lf}), yielding
\begin{equation}
\label{eq:sumrule_fp_sless2}
\Sigma_{\rm fp}(s) = 1-\frac{s}{2}\qquad\text{for $0<s<2$}\,.
\end{equation}
In the low-temperature limit, making use of (\ref{eq:c_inf_0}) we thus obtain
\cite{spren13}
\begin{equation}
\label{eq:c0_fp}
C_0 = \frac{s-1}{2}\qquad\text{for $0<s<2$}\,,
\end{equation}
which can only be valid as long as the size of the spatial region to which
the damped particle is confined is irrelevant. We conclude from (\ref{eq:c0_fp}),
that in the absence of finite-size effects the specific heat approaches zero
only for the special case of Ohmic damping. In the second regime, $s\geq 2$, 
one obtains 
\begin{equation}\label{eq:sumrule_fp_sgr2}
\Sigma_{\rm fp}(s)=0\qquad\text{for $s\ge 2$}\,.
\end{equation}
Thus, for $s\ge 2$ the specific heat tends to its classical value at very low
temperatures. This phenomenon has been termed ``reentrant classicality''
\cite{spren13}. Clearly, the expressions (\ref{eq:sumrule_osc}),
(\ref{eq:sumrule_fp_sless2}), and  (\ref{eq:sumrule_fp_sgr2}) are in
correspondence with the former result (\ref{eq:sum_diff_s}).

In order to express the specific heat of the damped free particle for arbitrary
values of $s$, (\ref{eq:c_xi_infty}) is most convenient and we get
\cite{spren13}
\begin{equation}
C = \frac{k_\text{B}}{2}+\int_0^\infty\text{d}\omega\xi_\text{fp}(\omega)
[C_\text{ho}(\omega)-k_\text{B}]\,.
\end{equation} 

\begin{figure}
 \centerline{\includegraphics[width=0.95\columnwidth]{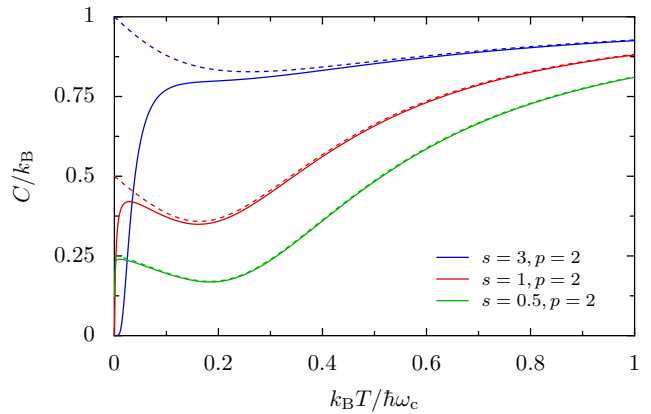}}
 \caption{Specific heat of a damped harmonic oscillator with
          $\omega_0=0.2\omega_\text{c}$ (solid line) and a free damped particle
          (dashed line). For ease of comparison, $k_\text{B}/2$ has been added
          to the specific heat of the free damped particle, i.e., the respective
          curves have been shifted upwards. The three sets correspond to the
          bath parameters $s=3, 1,$ and $0.5$ and $p=2$ from the upper to the
          lower set. The choice of $\gamma=5\omega_\text{c}$ implies that the
          specific heat of the free damped particle in the Ohmic case, $s=1$,
          becomes negative at low temperatures.}
 \label{fig:specificheat}
\end{figure}

Together with the results presented in Sect.~\ref{sec:change_of_bath_dos}, we
can now discuss how the thermodynamic anomalies of the free damped particle
manifest themselves for the damped harmonic oscillator.  We will mainly focus
on the Ohmic environment ($s=1$) and a superohmic environment with $s=3$ for
which the changes of the bath density of states have been shown in
Figs.~\ref{fig:changedos_1_2} and \ref{fig:changedos_3_2}, respectively.
Figure~\ref{fig:specificheat} displays the corresponding specific heats for a
damped harmonic oscillator with frequency $\omega_0=0.2\omega_\text{c}$ (full
line) and a free damped particle (dashed line). The middle and upper set of
curves correspond to $s=1$ and $s=3$, respectively, and $p=2$.  The damping
constant $\gamma=5\omega_\text{c}$ was chosen sufficiently large that the
specific heat for the free particle becomes negative at low temperatures
\cite{spren13}. The lower set of curves corresponds to $s=0.5$ and $p=2$ and
will be discussed at the end of this section.  Note that for better comparison
between the specific heat of the harmonic oscillator and the free particle, we
have shifted the specific heat of the latter by $k_\text{B}/2$, so that all
curves reach a value of $k_\text{B}$ in the high-temperature limit.

Comparing the curves for the harmonic oscillator with the shifted ones for the
free particle, we find that they agree rather well as long as the temperature
does not become too low. This behavior is consistent with what we had found
for the change of the bath density of states in Figs.~\ref{fig:changedos_1_2}
and \ref{fig:changedos_3_2}. The important difference there was the presence
of a low-frequency peak in the case of the harmonic oscillator. Its weight
accounts for the difference of $k_\text{B}/2$ in the classical values for the
specific heat for harmonic oscillator and free particle. The peak is also
responsible for the dramatic increase of the specific heat at low temperatures
visible in Fig.~\ref{fig:specificheat}.

We now take a closer look at how the thermodynamic anomalies for the free
damped particle appear in the specific heat of the damped harmonic oscillator.
We start with the middle set of curves displayed in Fig.~\ref{fig:specificheat}
corresponding to Ohmic damping. Since the dashed curve has been shifted upwards
by $k_\text{B}/2$, both curves actually start at vanishing specific heat. The
dashed curve thus exhibits the phenomenon of negative specific heat for the
free damped particle. In the case of the damped harmonic oscillator, the 
Lorentzian centered at zero frequency shown in Fig.~\ref{fig:changedos_1_2}
increases the specific heat by almost $k_\text{B}/2$ at rather low temperatures.
Then, the negative part of the change of the bath density of states shown
in Fig.~\ref{fig:changedos_1_2} leads to a decrease of the specific heat
with increasing temperature resulting in a pronounced dip. At higher temperatures,
the classical value for the specific heat of $k_\text{B}$ is approached from
below. We can thus conclude that while for the damped harmonic oscillator,
the specific heat cannot become negative, a dip may appear which has the
same physical origin as the negative specific heat for the free damped
particle. Such a dip in the specific heat can even be found if the heat bath
is constituted by a single harmonic oscillator. Its origin was traced back
to a frequency shift due to the coupling between system and bath \cite{ingol09}.
The approach in Sect.~\ref{sec:change_of_bath_dos} provides a generalization
from a one-oscillator bath to a bath consisting of many harmonic oscillators.

\begin{figure}
 \centerline{\includegraphics[width=0.95\columnwidth]{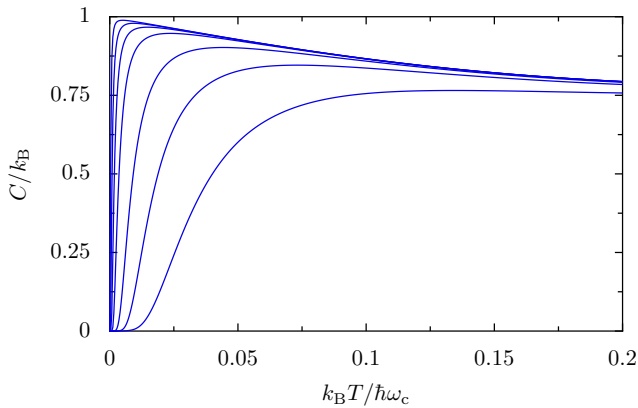}}
 \caption{Building-up of reentrant behavior for a damped harmonic oscillator 
          with frequency $\omega_0/\omega_\text{c}=0.2, 0.1, 0.05, 0.02, 0.01,
          0.005$, and $0.002$ decreasing from the lower to the upper curve. The
          damping strength is $\gamma=5\omega_\text{c}$ and the bath density of
          states is characterized by $s=3$ and $p=2$.}
 \label{fig:specificheat_3_2}
\end{figure}

Let us now turn to the upper set of curves in Fig.~\ref{fig:specificheat}
corresponding to a superohmic environment with $s=3$. The dashed curve again
displays the specific heat of the free damped particle shifted upwards by
$k_\text{B}/2$. As we can see, the specific heat exhibits reentrant classicality
in the sense that the reduction of the specific heat at intermediate temperatures
disappears at low temperatures where the classical high-temperature value is
approached. As already discussed, the specific heat for the damped harmonic
oscillator coincides with the one of the free particle up to a shift of 
$k_\text{B}/2$ for sufficiently large temperatures. At lower temperatures,
deviations necessarily occur because the specific heat of the damped harmonic
oscillator has to approach zero in the zero-temperature limit. In order to
verify that in the limit of vanishing oscillator frequency $\omega_0\to0$
reentrant classicality does occur, we need to consider the specific heat
for smaller values of $\omega_0$ as depicted in Fig.~\ref{fig:specificheat_3_2}.
It can be seen that with decreasing oscillator frequency $\omega_0$, the
temperature range in which the specific heat drops from a value close to
$k_\text{B}$ down to zero becomes arbitrarily small. In this sense, reentrant
classicality can be observed even for a damped harmonic oscillator.

We close this section by taking a look at the lower set of curves of
Fig.~\ref{fig:specificheat} corresponding to $s=0.5$ and $p=2$. While these
curves in the view of the previous discussion provide no surprises, the curve
for the damped harmonic oscillator supports the result found earlier for the
free damped particle with $s<1$ that the specific heat tends towards a negative
value in the low-temperature limit. Keeping in mind the offset of
$k_\text{B}/2$, for $s=0.5$, a value of $C=-0.25k_\text{B}$ is indeed approached
for small temperatures. Very much like the curve for the damped harmonic
oscillator drops down to a vanishing specific heat at zero temperature, one
would expect that accounting for finite-size effects would lead to a sharp
increase of the specific heat of the damped free particle as zero temperature is
approached.

\section{Entropy}
\label{sec:entropy}

Another thermodynamic quantity of interest is the entropy. A nonmonotonic
temperature dependence of entropy has been discussed e.g. for Kondo systems
\cite{zitko09,flore04}, XY spin chains \cite{campi10}, and two-level fluctuators
\cite{campi09}. In the context of the Casimir effect, the appearance
of a negative entropy due to a finite zero-frequency electric conductivity has
been the subject of an extensive debate
\cite{klimc06,hoye07,milto09,ingol09a,intra09}. A relation between this problem
and the damped free particle was pointed out \cite{ingol09a} and the
temperature dependence of the entropy of a damped free particle bears
resemblance with curves obtained for the Casimir entropy for real metals
\cite{bostr04}.

In comparison with the specific heat, putting free damped particle and damped
harmonic oscillator into relation is complicated by the fact that the classical
expressions for the entropy differ by a function of temperature. For the damped
harmonic oscillator, one obtains from the first factor in (\ref{eq:z_ho})
\begin{equation}
\label{eq:shocl}
S_\text{osc,cl} =
k_\text{B}\left[\ln\left(\frac{k_\text{B}T}{\hbar\omega_0}\right)+1\right]\,.
\end{equation}
For the free damped particle, one obtains instead
\begin{equation}
\label{eq:sfpcl}
S_\text{fp,cl} =
S_0+\frac{k_\text{B}}{2}\left[\ln\left(\frac{k_\text{B}T}{\hbar\gamma}\right)+1
\right]
\end{equation}
where
\begin{equation}
\label{eq:s0}
S_0 =
\frac{k_\text{B}}{2}\ln\left(\frac{\pi}{4}\frac{\hbar\gamma}{E_\text{g}}\right)\,.
\end{equation}
It can be shown that $S_0$ is the zero-temperature value of the entropy of the
free damped particle if finite-size effects are disregarded \cite{ingol09a}.

For the discussion of the entropy, we focus on an Ohmic heat bath with standard
Drude cutoff, i.e.\ $s=1$ and $p=1$. In this special case, the reduced partition
functions (\ref{eq:z_fp}) and (\ref{eq:z_ho}) can be expressed as
\begin{equation}
\mathcal{Z} = \mathcal{Z}_\text{cl}\dfrac{\prod_n\Gamma\left(1-\dfrac{\hbar\beta z_n^0}
{2\pi}\right)}{\Gamma\left(1+\dfrac{\hbar\beta\omega_\text{c}}{2\pi}\right)}
\end{equation}
where $\Gamma(x)$ is the gamma function. $\mathcal{Z}_\text{cl}$ is the
classical contribution to the reduced partition function, i.e.\ the factor in
front of the infinite product in (\ref{eq:z_fp}) and (\ref{eq:z_ho}).  The
quantities $z_n^0$ are obtained as zeros of the polynomial
\begin{equation}
N(z;\omega_0) = z^3+z^2\omega_\text{c}+z(\gamma\omega_\text{c}+\omega_0^2)
+\omega_0^2\omega_\text{c}
\end{equation}
for the damped harmonic oscillator and
\begin{equation}
\frac{1}{z}N(z;0) = z^2+z\omega_\text{c}+\gamma\omega_\text{c}
\end{equation}
for the damped free particle.

Using standard
relations from thermodynamics, the entropy can then be expressed as
\begin{equation}
S =
S_\text{cl}+k_\text{B}\left[-f\left(\frac{\hbar\beta\omega_\text{c}}{2\pi}\right)
+\sum_n f\left(-\frac{\hbar\beta z_n^0}{2\pi}\right)\right]\,,
\end{equation}
where $S_\text{cl}$ for the harmonic oscillator and free particle is given by
(\ref{eq:shocl}) and (\ref{eq:sfpcl}), respectively. Furthermore,
\begin{equation}
f(x) = \ln[\Gamma(1+x)]-x\psi(1+x)
\end{equation}
with the digamma function $\psi(x)$.

\begin{figure}
 \centerline{\includegraphics[width=0.95\columnwidth]{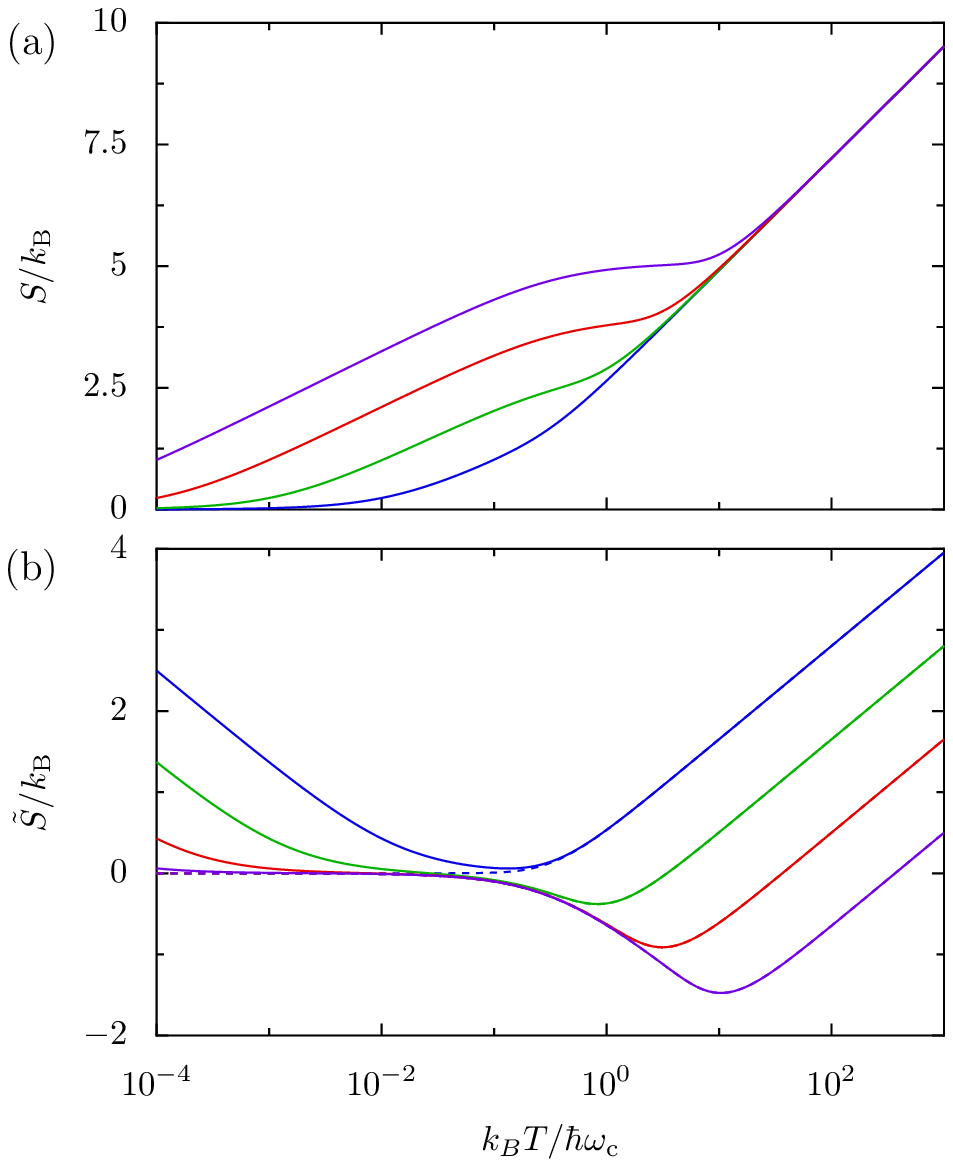}}
 \caption{(a) Entropy of a damped harmonic oscillator with $\omega_0=0.2\omega_\text{c}$
          coupled to an Ohmic environment ($s=1$) with Drude cutoff ($p=1$) and
          varying damping strength $\gamma/\omega_\text{c}=1, 10, 100,$ and $1000$ from 
          the lower to the upper curve. (b) The solid lines represent the entropy
          $\tilde S$ defined in (\ref{eq:stilde}) and obtained from the data
          represented in panel (a) by means of (\ref{eq:shocl}) and
          (\ref{eq:sfpcl}). The values of the damping strengths coincide with
          those chosen for panel (a), but here they increase from the upper to
          the lower curve. The dashed lines represent the entropy $S-S_0$ of the free
          damped particle disregarding finite-size effects. Because the entropy
          constant (\ref{eq:s0}) has been subtracted off all data in this panel,
          the dashed lines approach zero in the limit of vanishing temperature.}
 \label{fig:entropy}
\end{figure}

Figure~\ref{fig:entropy}(a) displays the entropy for a damped harmonic
oscillator with frequency $\omega_0=0.2\omega_\text{c}$ and damping strengths
$\gamma/\omega_\text{c}=1, 10, 100,$ and $1000$ increasing from the lower to
the upper curve. The straight-line behavior for large temperatures on this
semilogarithmic plot reflects the classical behavior (\ref{eq:shocl}) of the
entropy. For sufficiently strong damping, a crossover to a logarithmic behavior
characteristic of the classical free particle is observed as the temperature is
lowered. At even smaller temperatures, the oscillator frequency becomes
relevant again to ensure that the entropy goes to zero in accordance with the
third law of thermodynamics as temperature goes to zero.

Correcting for the difference in the classical behavior between harmonic
oscillator and free particle, as explained in the context of the reduced
partition function in Sect.~\ref{sec:reduced_partition_function}, we can
define an entropy
\begin{equation}
\label{eq:stilde}
\tilde S = S_\text{osc}-S_\text{osc,cl}+S_\text{fp,cl}
\end{equation}
which should agree with the entropy of the damped free particle except for not
too low temperatures. This correction for the classical behavior corresponds to
the shift by $k_\text{B}/2$ which we had applied to the specific heat of the
free damped particle in Fig.~\ref{fig:specificheat}. In
Fig.~\ref{fig:entropy}(b) we have taken the entropy data displayed in the upper
panel to obtain together with (\ref{eq:shocl}) and (\ref{eq:sfpcl}) the entropy
$\tilde S$ defined in (\ref{eq:stilde}) represented as solid lines. The damping
strength increases here from the upper to the lower curve. For comparison, the
dashed lines show the results for the entropy of the free damped particle. The
larger the damping strength, the longer $\tilde S$ agrees with the entropy
$S-S_0$ of the free particle as the temperature is lowered. Only for even lower
temperatures, the finite frequency of the oscillator used in determining
$\tilde S$ becomes relevant.  Note that in the curves shown in
Fig.~\ref{fig:entropy}(b), the constant (\ref{eq:s0}) has been subtracted. Its
dependence on the damping strength is responsible for the reversed order of the
curves in Figs.~\ref{fig:entropy}(a) and (b).

It was already shown in Ref.~\cite{ingol09a} that the entropy of the damped
free particle at finite temperature can be smaller than its zero-temperature
value as long as finite-size effects are disregarded. As we can see from
Fig.~\ref{fig:entropy}, this effect is visible in the temperature dependence
of the entropy of the damped harmonic oscillator in the form of a plateau
separating the regime dominated by the classical behavior and the one dominated
by the free particle classical behavior. This thermodynamic scenario complements
the scenario for the dynamics of an overdamped harmonic oscillator where free
particle behavior is observed for not too long times \cite{jung85}. 

The results based on the entropy of the damped harmonic oscillator shown as
solid lines in Fig.~\ref{fig:entropy}(b) indicate how the entropy of a free
damped particle constrained to a finite-size region behaves as a function of
temperature. The entropy can indeed be a nonmonotonic function of the
temperature. However, as the deviations between $\tilde S$ and the entropy of the
free particle show, there will be a potentially very small temperature interval
on which the entropy changes its value from $S_0$ to zero as required by the
third law of thermodynamics when zero temperature is approached.

\section{Conclusions}
\label{sec:conclusions}

Thermodynamic quantities of a damped free particle cannot directly be obtained
from those of the damped harmonic oscillator by taking the limit of vanishing
oscillator frequency. Nevertheless, it is possible to identify correspondences
between thermodynamic anomalies of the damped free particle and features in
the temperature dependence of the respective quantities for the damped harmonic
oscillator. Negative specific heats for the free damped particle correspond to
dips in the specific heat of the damped harmonic oscillator and minima of the
entropy of the free damped particle can be related to plateaus in the entropy
of the damped harmonic oscillator. Reentrant classicality can be found both
for the damped free particle and the damped harmonic oscillator for superohmic
environments with $s\geq 2$.

From the bath point of view, the main difference between the two damped systems
consists in an additional peak in the change of the bath density of states
either at zero frequency or at a low but finite frequency. This peak is
relevant for the behavior of thermodynamic quantities in the classical regime
and continues to describe the difference between damped free particle and
damped harmonic oscillator down to low temperatures before finite-size effects
set in.

The results for the damped harmonic oscillator show that the transition to
vanishing specific heat and entropy as temperature goes to zero can occur
within a rather small temperature range. It can be expected that the
regularization of the thermodynamic quantities of a free damped particle due to
confinement to a finite-size region behaves in a very similar way to ensure the
validity of the third law of thermodynamics.

\begin{acknowledgement}
One of us (GLI) would like to thank Peter H{\"a}nggi, Peter Talkner, Astrid
Lambrecht, and Serge Reynaud for useful discussions. He also gratefully
acknowledges the hospitality of the Laboratoire Kastler Brossel in Paris during
the preparation of this manuscript. UW has received financial support from
the Deutsche Forschungsgemeinschaft through SFB/TRR 21.
\end{acknowledgement}

\appendix

\section{Low-frequency peak in the change of the bath density of states
for arbitrary \boldmath $s$}
\label{sec:appa}

In Sect.~\ref{sec:change_of_bath_dos} we have restricted the discussion of
the low-frequency peak in the change of the bath density of states to the
Ohmic case, $s=1$, as well as the superohmic case $s=3$. In this appendix,
we provide the generalization to arbitrary values of $s$.

To obtain the low-frequency peak of $\xi_{\rm osc}(\omega)$ for non-integer
values of $s$, we assume $\omega \ll \omega_{\rm c}$ so that we may truncate
the expression (\ref{eq:gammahat_int_p}) to the form
\begin{equation}
\label{eq:gamma_lf}
\hat\gamma(z) =\frac{ \gamma}{\sin(\frac{\pi s}{2})}
  \left(\frac{z}{\omega_{\rm c}}\right)^{s-1}  + \mu z\,. 
\end{equation}
Here, $\mu =\Delta M/M$ is the relative mass contribution of the bath
\cite{spren13}
\begin{equation}
\mu(s,p) = \frac{2}{\pi} \frac{\gamma}{\omega_{\rm c}}
\frac{p B(\frac{s}{2},p+1-\frac{s}{2})}{s-2} \, .
\end{equation}
The first term in (\ref{eq:gamma_lf}) has a branch cut at $z=0$.

In the regime $0 < s <2$ and $\omega\ll \omega_{\rm c}$, the first term in
(\ref{eq:gamma_lf}) is the leading one. A low-frequency approximation of
the dynamical susceptibility of the damped harmonic oscillator and the
dynamical velocity response of the damped free particle is obtained by
inserting this leading term into (\ref{eq:chi_osc}) and (\ref{eq:R_fp}),
respectively. By means of the expressions (\ref{eq:xi_osc}) and
(\ref{eq:xi_fp}), we find that $\xi_{\rm osc}(\omega)$ and $\xi_{\rm
fp}(\omega)$ differ by a low-frequency peak of the form
\begin{equation}
\label{eq:lfpeak1}
\xi_{\rm osc}(\omega)-\xi_{\rm fp}(\omega) = \frac{s}{\pi}
\left(\frac{\omega}{\omega_{\rm c}}\right)^{s-1}  
\!\!\! \frac{\Gamma_\text{r}}{\Gamma_\text{r}^2 + \left(\dfrac{\omega^s}{\omega_\text{c}^{s-1}} +
\Omega_\text{r}\right)^2}
\end{equation}
with the parameters
\begin{equation}\label{eq:para1}
\begin{aligned}
\Gamma_{\rm r} &=\frac{\omega_0^2}{\gamma}\sin\left(\frac{\pi s}{2}\right)^2\,,\\ 
\Omega_{\rm r} &=\frac{\omega_0^2}{2\gamma}\sin(\pi s)\,.
\end{aligned}
\end{equation}
The expression (\ref{eq:lfpeak1}) diverges at $\omega=0$ for subohmic damping,
$s<1$, while it goes to zero and displays a maximum at finite frequency in the
superohmic regime $1<s<2$. The two regimes adjoin at the Ohmic point $s=1$
where $\xi_{\rm osc}(0) - \xi_{\rm fp}(0)=\gamma/\pi\omega_0^2$. For the
parameters used in Fig.~\ref{fig:changedos_1_2} the peak reaches
$39.8/\omega_\text{c}$ and is too high to be shown.

The expression (\ref{eq:lfpeak1}) with (\ref{eq:para1}) is a reasonable
approximation in the regime $0<s \lesssim 1.5$. As $s$ approaches 2, cut-off
dependent contributions, which are absent in (\ref{eq:para1}), become
increasingly relevant in the expressions for $\Gamma_{\rm r}$ and
$\Omega_{\rm r}$.

In the superohmic regime $s>2$ also the second term in (\ref{eq:gamma_lf})
is relevant. Using the form (\ref{eq:gamma_lf}) for $\hat\gamma(z)$, the
low-frequency behavior of the density is found as
\begin{equation}
\label{eq:lfpeak2}
\xi_{\rm osc}(\omega) -\xi_{\rm fp}(\omega) = \frac{1}{\pi} \sum_{r=\pm 1} \frac{\Gamma}{
(\omega- r \Omega)^2+\Gamma^2} \; .
\end{equation}
The parameters $\Gamma$ and $\Omega$ read
\begin{equation}
\label{eq:para2}
\begin{aligned}
\Gamma &=  \frac{\gamma}{2}
\frac{ \left(\omega_0/\omega_{\rm c}\right)^{s-1} }{(1+ \mu)^{(1+s)/2} }\\
\Omega &= \frac{\omega_0}{\sqrt{1+\mu}}\,,
\end{aligned}
\end{equation}
where terms of order $(\omega_0/\omega_{\rm c})^2$ and $(\omega_0/\omega_{\rm
c})^{s-2}$ have been omitted. As $s$ approaches 2 from above, these terms
become increasingly important. The expressions (\ref{eq:para2}) effectively
apply in the regime $s\gtrsim 2.5$.

From (\ref{eq:lfpeak1}) and  (\ref{eq:lfpeak2}), we find as an important
result the expressions given for $\Delta\Sigma(s)$ in (\ref{eq:sum_diff_s}).

\end{document}